\documentclass[preprintnumbers,amsmath,amssymb,prd,notitlepage, twocolumn,nofootinbib]{revtex4-1}
\pdfoutput=1
\usepackage[english]{babel}
\usepackage[utf8]{inputenc}
\usepackage{graphicx}
\usepackage{subfigure}
\graphicspath{ {slike/} }

\newcommand{\be}{\begin{equation}}
\newcommand{\ee}{\end{equation}}

\newcommand{\lya}{Ly$\alpha$}
\newcommand{\lyb}{Ly$\beta$}

\begin{document}

\title{Background power subtraction in \lya\ forest}

\author{Vid Ir\v{s}i\v{c}}
\email{virsic@ictp.it}
\affiliation{The Abdus Salam International Center for Theoretical
  Physics, Strada Costiera, 11, Trieste 34151, Italy}
\author{An\v{z}e Slosar}
\email{anze@bnl.gov}
\affiliation{Brookhaven National Laboratory, Bldg 510, Upton NY 11375, USA}

\date{\today}

\begin{abstract}
  When measuring the one-dimensional power spectrum of the \lya\
  forest, it is common to measure the power spectrum in the flux
  fluctuations red-ward of the \lya\ emission of quasars and subtract
  this power from the measurements of the \lya\ flux power
  spectrum. This removes the excess power present in the \lya\ forest
  which is believed to be dominated by the metal absorption by the
  low-redshift metals uncorrelated with the neutral hydrogen absorbing
  in \lya. In this brief report we note that, assuming the
  contaminants are additive in the optical depth, the correction
  contains a second order term. We estimate the magnitude of this term
  for two currently published measurements of the 1D \lya\ flux power
  spectrum and show that it is negligible for the current generation
  of measurements. However, future measurements will have to take this
  effect into account when the errorbars improve by a factor of two or
  more.
\end{abstract}

\maketitle

\section{Introduction}

The Lyman-$\alpha$ forest measurements are becoming increasingly more
accurate and to that end careful investigation of possible systematic
effects is required. In this brief report we study the effect of the
background power fluctuations which contaminate the signal in the
\lya\ forest region.

Fluctuations in the \lya\ forest region in the spectra of distant
quasars, that is region between the rest \lya\ and \lyb\ emission
lines (with some buffer to immunize against proximity effects) is
dominated by the \lya\ absorption. However, metals in the
inter-galactic medium will contaminate this signal coming from neutral
hydrogen. There are several techniques to attack this important
systematics.  For metal transitions which occur at wavelengths similar
to the \lya\ emission wavelength ($\lambda_\alpha = 1215.67$\AA) we
can rely on the fact that the contaminant metals are closely tracing
the dominant absorption by neutral hydrogen producing detectable
``beating'' in the power spectrum measurements.  This has been
demonstrated in \cite{2006ApJS..163...80M,2013A&A...559A..85P} for Si
III and \cite{2013JCAP...09..016I} for O VI. On the other hand it is
relatively easy to remove contribution to absorption by metals whose
transitions $\lambda$ are sufficiently larger than
$\lambda_\alpha$. The most common way to do this is to measure the
power spectrum of fluctuations redward of the \lya\ emission in the
spectra of quasars. Since gas behind a given quasar cannot absorb
quasar light, this power is often termed the background power -- the
power spectrum in absence of signal -- and subtracted from the
measured flux power spectrum. 

Note that for a given observed wavelength $\lambda_o$, there are
quasars at somewhat larger redshift $1+z>\lambda_o/\lambda_\alpha$ for
which the wavelength is subject to both \lya\ and contaminant
absorptions and \emph{other} quasars at somewhat lower redshifts
$1+z<\lambda_o/\lambda_\alpha$ for which the same wavelength is
absorbed only by the contaminant. Therefore, we are correcting the
observed \lya\ forest flux power spectrum by subtracting contaminant
flux power measured in the lower redshift quasars, but corresponding
to the same observed wavelength range and thus to the statistically
the same component.
 
It is believed that majority of the contaminant signal is coming from
a mixture of metal absorptions by a lower redshift ($z<1.5$)
intergalactic medium. However, this simple subtraction will remove all
absorption associated with metal lines with rest-frame wavelength
falling red-ward of the region in which contaminant power is
estimated. It is most common to use the region $1270<\lambda<1380$\AA,
which also removes significant  absorption due to both Si IV (a
doublet at rest wavelengths 1393.75\AA and 1402.77\AA)
and C IV (another doublet absorbing at 1548.20\AA and 1550.78\AA).

Since the gas casing contaminant absorption is physically very far
from the gas causing primary \lya\ forest, the fluctuations in the two
are uncorrelated.  However, the contaminant signal adds to the total
optical depth experienced by the quasar's photons, which leads to a
second order effect, which we discuss in this work.  We will show that
this effect is negligible for the present generation of the
one-dimensional flux power spectra measurements, but that it will
likely become important for the final BOSS
(\cite{2013AJ....145...10D,2012AJ....144..144B}) analysis, eBOSS
(\cite{eBOSS}) and DESI (\cite{2013arXiv1308.0847L}) experiments.

\section{Effect of contaminants}

For our analysis, we assume that the observed flux in the relevant parts
of quasar spectra is given by 
\begin{multline}
f^q(\lambda_i) = C^q(\lambda_r)
\times \begin{cases}
 {e^{-\tau_\alpha(z_i)-\tau_c(z_i)}} & \mbox{in forest region}\\
 {e^{-\tau_c(z_i)}} & \mbox{in background region}\\
\end{cases},
\end{multline}
where $C^q(\lambda_r)$ is the continuum of the quasar (results in this
paper do not depend on the modeling of this quantity) and
$\tau_\alpha$ and $\tau_z$ are the optical depths associated with
the signal and the contamination respectively. We write the absorptions as
\begin{eqnarray}
 e^{-\tau_\alpha(z_i)} &=& {\bar F}_\alpha (z_i)  (1 +
 \delta_\alpha(\lambda_i) ) \\
 e^{-\tau_c(z_i)} &=& {\bar F}_c (z_i)  (1 + \delta_c(\lambda_i) )
\end{eqnarray}
where $\bar{F}$s are the mean absorptions and $\delta$s are the
corresponding fluctuations for the two components. Here we have
written $\delta_c$ as fluctuations due to the low-redhift metals, but
fluctuations due to the continuum errors would have exactly the same
form (i.e. be multiplicative).  The forest region of the spectrum is
the region of the forest blue-ward of \lya\ emission, typically
between the rest frame wavelengths of 1050\AA\ and 1185\AA. This range is
chosen to avoid proximity effects near the emission lines of both
\lya\ and \lyb\ where behavior of the continuum is poorly understood.
Different authors have defined different background regions where the
absorption by metals is measured but all regions share the same
qualities of having being as close as possible redward of \lya\
emission and having a smooth continuum without large emission lines (see
\cite{2006ApJS..163...80M,2013A&A...559A..85P,2013A&A...552A..96B}).

Faced with the real quasar spectra, it is impossible to distinguish
between the various absorbers and the best one can hope to do is to
model the quasar flux as
\cite{2006ApJS..163...80M,2013A&A...559A..85P,2013JCAP...04..026S,2013A&A...552A..96B}
\begin{multline}
f^q(\lambda_i) = C^q(\lambda_r)\\
\times \begin{cases}
 {\bar F}_T (z_i) (1 + \delta_T (\lambda_i)) & \mbox{in the forest region}\\
 {\bar F}_c(z_i) (1 + \delta_c(\lambda_i)) & \mbox{in the redward region}\\
\end{cases},
\label{eq:1}
\end{multline}
where subscript $T$ stands for total absorption of both \lya\ and
contaminant.
Inside the forest region the total fluctuation field
can be written as 
\be 1 + \delta_T(\lambda) = (1 +
\delta_\alpha(\lambda)) (1 + \delta_c(\lambda)).
\label{eq:2}
\ee
We see that in addition to the usual linear relation
used in the previous work, there is also a second order (quadratic) term
in the total absorption flux fluctuation field in the forest
\be
\delta_T(\lambda) = \delta_\alpha(\lambda) + \delta_c(\lambda) +
\delta_\alpha(\lambda)\delta_c(\lambda).
\label{eq:3}
\ee
The correlation function of this quantity is given by
\begin{multline}
\langle \delta_T (\lambda) \delta_T(\lambda') \rangle =
\xi_{T}(\lambda, \lambda')= \\
\xi_{\alpha}(\lambda, \lambda') +
\xi_{c}(\lambda, \lambda')+
\xi_{\alpha}(\lambda, \lambda') \xi_{c} (\lambda, \lambda')
\label{eq:5}
\end{multline}

Here we have assumed that $\delta_\alpha$ and $\delta_c$ are
completely uncorrelated fields. This is justified by the fact that the
metals causing the contaminant absorption are sitting in a gas that is
typically $>500\mathrm{h^{-1}Mpc}$ away from the hydrogen gas
absorbing in \lya.

\begin{figure*}
  \centering
  \includegraphics[width=1.0\linewidth]{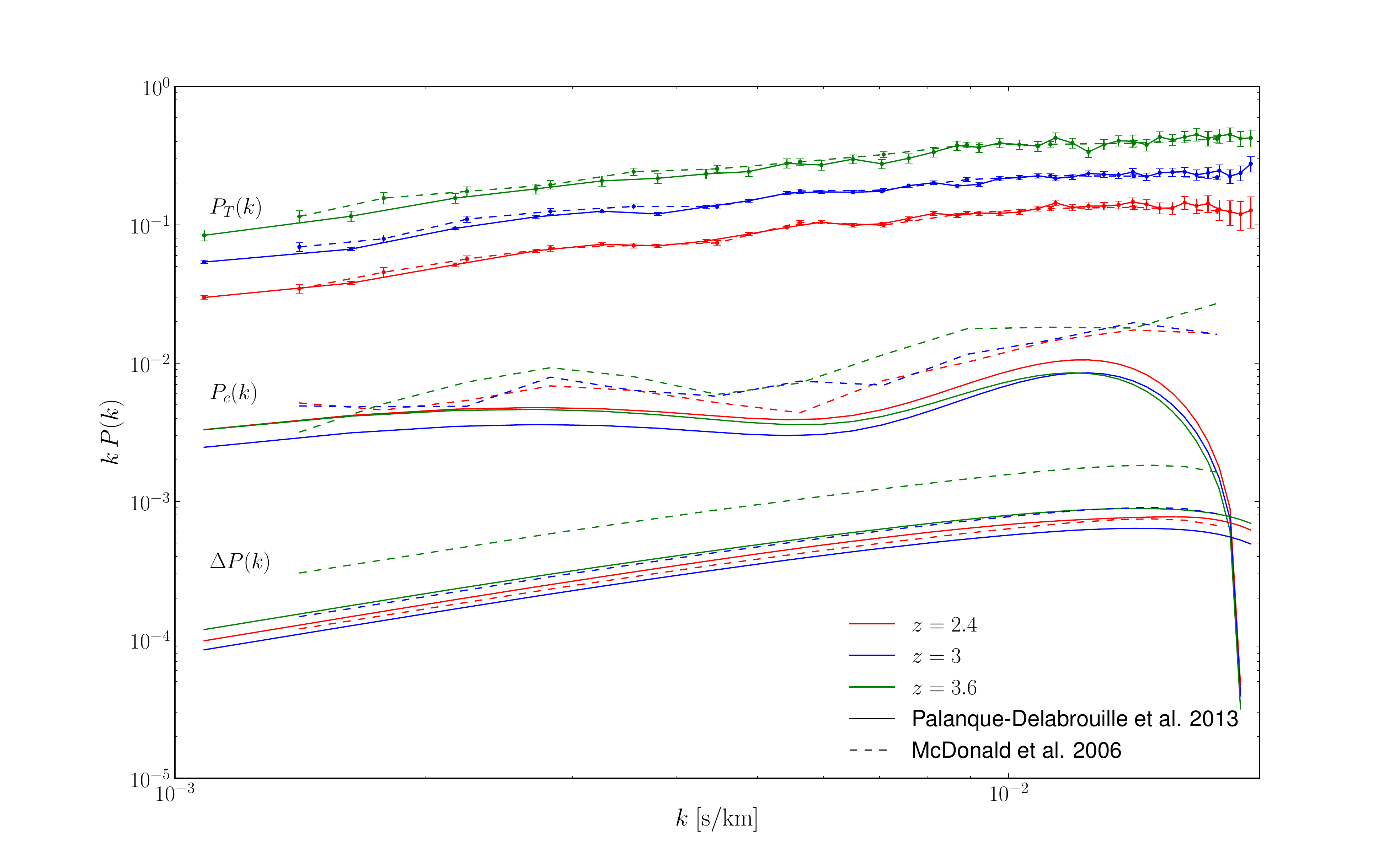}
  \caption{
    Total power spectrum, contaminant power spectrum and second order correction
    for 3 redshifts. Different colors correspond to different
    redshift bins, while different line-styles correspond to different
    analyses of the power spectra as denoted in the legend. See text
    for further discussion.
  }
  \label{Fig:1}
\end{figure*}

\begin{figure}
  \centering
  \includegraphics[width=1.0\linewidth]{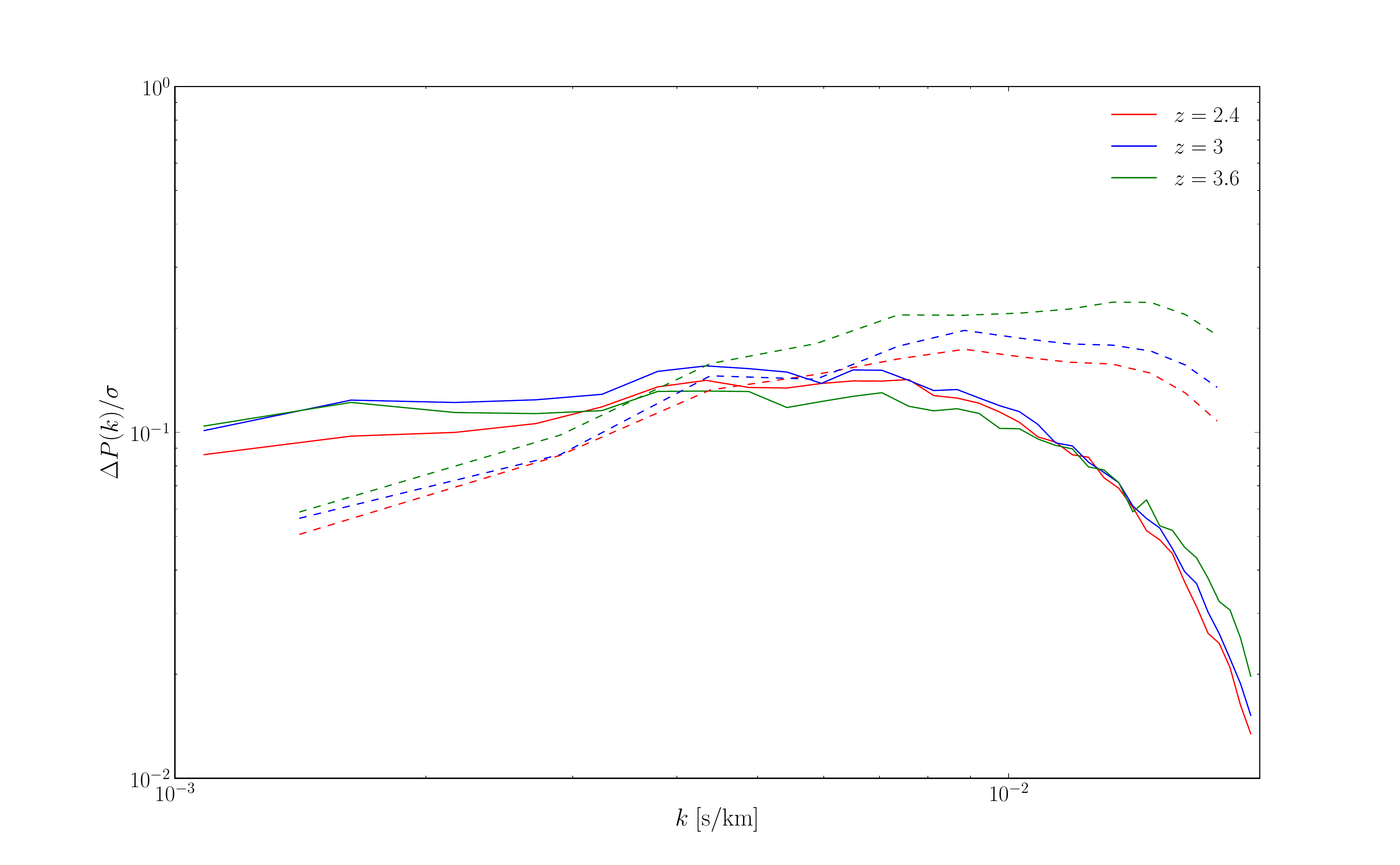}
  \caption{
    Correction with respect to the errors using the same color scheme
    as in Figure \ref{Fig:1}.
  }
  \label{Fig:2}
\end{figure}

Fourier transforming, we find that the power spectra are given by 
\begin{multline}
P_{T}(k) =  P_{\alpha}(k) + P_{c}(k) + \frac{1}{2\pi}\int_{-\infty}^\infty P_{\alpha}(k')P_c(k-k') dk'
\\
= P_{\alpha}(k) + P_{c}(k)+(P_{\alpha}\star P_{c})(k),
\label{eq:6}
\end{multline}
where $\star$ stands for convolution.

On the quasar's red-side, there is no forest and therefore the
measured power spectrum is simply $P_{c}(k)$. The point of this short
note is that when a corrected power spectrum is calculated, the
second order correction does not cancel 
\begin{multline}
  P_{\rm std. correction}(k) = P_{T}(k) - P_{c}(k)\\ = P_{\alpha}(k) 
  + (P_{\alpha}\star P_{c})(k).
\end{multline}

Therefore, to recover the signal power spectrum $P_\alpha$ it is not
sufficient to simply subtract the contaminant power spectrum from the
total power spectrum.  In configuration space
\be
\xi_{\alpha}(x) = \frac{\xi_{T}(x) -
  \xi_{c}(x)}{1 + \xi_{c}(x)},
\ee
which gives
\be
P_{\alpha}(k) = P_{T}(k) - P_{c}(k) - \Delta P(k),
\label{eq:9}
\ee
where
\be
\Delta P(k) = \int_{-\infty}^\infty dk' \left[ P_{T}(k') - P_{c}(k')
\right] {\tilde W}(k-k'),\\
\label{eq:10}
\ee
and ${\tilde W}$ is just the Fourier transform of the real space window
function
\be
W(x) = \frac{\xi_{c}(x)}{1 + \xi_{c}(x)},
\label{eq:12}
\ee

There are two interesting limits to these equations. First, if
$\xi_c(x)$ is small compared to other quantities, we see that
${\tilde W}(k)\sim P_c(k)$, leading to 
\begin{equation}
  \Delta P (k) = (P_{\rm std. correction} \star P_c)(k)
\end{equation}
(in effect approximating $P_\alpha(k)$ with $P_{\rm std. correction}(k)$ in
Eq. \ref{eq:6}). Second, when $P_c(k)$ is white (i.e. $\xi_c(x) = \sigma_c^2\delta^D(x)$), we see that
\begin{equation}
  \Delta P(k)  = \frac{2\pi \sigma_c^2\sigma_\alpha^2}{k_{\rm max}},
\end{equation}
where $\sigma$s are the variances (i.e. zero lag correlators) of the
$\alpha$ and $c$ fields. We see that in that limit, the correction is
purely white too.

In order to properly account for this effect, one would need to take
it into account in the quadratic estimator used to measure the power
spectrum (and likely perform the measurements of the background and
the forest power jointly). However, to get a rough estimate, one can
take measurements of the power spectra, Fourier transform those
measurements to the configuration space, perform correction and
transform back. We do this for the two published results in the next
section.

\section{Estimating the size of the effect}

To evaluate the correction from Eq. \ref{eq:10} we have used the FFT
algorithm to first compute the corresponding $\xi_{c}$ and
$\xi_{\alpha}$ and then to compute the inverse as given by
Eq. \ref{eq:9}. We did this for two published 1D power spectra
\cite{2006ApJS..163...80M,2013A&A...559A..85P}, which conveniently
provided both the total power measured in the \lya\ forest region as
well as background power measured redward of the \lya\ emission
line. To deal with different binning schemes and the finite $k$-space
coverage, we have first re-sampled the power spectra onto a finer
$k$-space grid and zero-padded on both sides until results
converged. We have also checked that treating the power spectrum bins
as either flat bandpower bins or linearly interpolating between bin
centers made negligible difference.

In the Figure \ref{Fig:1} we plot the quantities $P_T$, $P_m$ and
$\Delta P$ on the same plot for three representative redshift
bins. This plot shows that the correction is small, three orders of
magnitude smaller than the power spectrum and an order of magnitude
smaller than the background power spectrum. The full lines show the
results from \cite{2013A&A...559A..85P}, while the dashed lines show
an earlier analysis by \cite{2006ApJS..163...80M}.

Note that due to small errorbars the correction is not as negligible
as one might naively expect. In the Figure \ref{Fig:2} we show the
size of the correction relative to the power spectra error estimates
(both statistical and systematic). We see that for the current
generation of power spectra measurements, the 2nd order correction is
not yet important (around $10\%$ of current errors in the redshift
range $z=3-4$).  The implied change in $\chi^2$ for previous works is
$\sim 1$ for \cite{2006ApJS..163...80M} (over $12$ $k$ bins and $11$
redshift bins) and $\sim 2$ for \cite{2013A&A...559A..85P} (over 35
$k$-bins and $12$ redshift bins). This confirms that the size of the
effect is probably negligible when compared to the current error-bars,
but not by a large margin.

\section{Conclusions}

In this brief report we have shown that the multiplicative
contaminations in the Lyman-$\alpha$ forest, such as those arising by
the intervening low redshift metals cannot be simply subtracted by
measuring them outside the forest region, but instead produce
higher-order corrections. These corrections are typically small and we
have demonstrated that they do not matter for the current generation
of the 1D power spectrum measurements.

The measurement of \cite{2013A&A...559A..85P} has used
approximately 14 thousands high signal-to-noise BOSS quasars,
producing an effect of $\Delta \chi^2\sim2$.  The
full survey will contain approximately 160 thousand quasars and eBOSS
and DESI experiments will likely increase the number of quasars to
well over 600 thousand. This signal to noise is hence likely to
increase by a factor of at least a few, bringing the expected size of
the effect well into the realm where correction will have to be applied.

Finally, we note that the correction mixes up small scales and large
scales. This can have important consequences, when continuum
fluctuations are taken into account. Traditionally, analyses have
relied on the fact that the continuum fluctuations, that is excess
fluctuations associated with the fact that un-absorbed continua vary
from quasar to quasar, are both slowly-varying with wavelength and
uncorrelated with the cosmic structure. Hence, the existing 1D power
spectrum measurements have limited their analyses to sufficiently
large wave-vectors (e.g. $k>10^{-3}$s/km) and the 3D analysis have
relied on cross-power.  Note, that we cannot simply subtract continuum
fluctuations from the red-side, because these are now in a wrong part
of the rest-frame spectrum: continuum fluctuations at 1100\AA\ are not
necessarily the same as continuum fluctuations at 1300\AA\
rest-frame. For standard correction, this does not matter, as we can
simply discard large scales. But the convolution in the second-order
correction discussed in this report in principle requires knowledge of
power at \emph{all} scales. This might set a fundamental limitation on
how well one can perform this correction, since the power on very
large and very small scales will most likely have to be estimated
using some form of extrapolation. We do not deal with this question in
this brief report, but undoubtedly new techniques will arrive that
will attack these issues.

\bibliographystyle{plain}
\bibliography{cosmo,cosmo_preprints}

\end{document}